\documentclass[aps,prl,twocolumn,amsmath,amssymb,showpacs,superscriptaddress]{revtex4-1}

\usepackage[colorlinks=true,linkcolor=blue,anchorcolor=red,citecolor=blue,urlcolor=blue]{hyperref}
\usepackage{bm}
\usepackage{graphicx}
\usepackage{color}
\usepackage{multirow}
\usepackage[final]{pdfpages}

\begin{document}

\title{Anomalous Phase Shift of Quantum Oscillations in 3D Topological Semimetals}

\author{C. M. Wang}

\affiliation{Department of Physics, South University of Science and Technology of China, Shenzhen 518055, China}
\affiliation{School of Physics and Electrical Engineering, Anyang Normal University, Anyang 455000, China}

\author{Hai-Zhou Lu }
\email{luhz@sustc.edu.cn}
\affiliation{Department of Physics, South University of Science and Technology of China, Shenzhen 518055, China}

\author{Shun-Qing Shen }
\affiliation{Department of Physics, The University of Hong Kong, Pokfulam Road, Hong Kong, China}

\date{\today}

\begin{abstract}
Berry phase physics is closely related to a number of topological states of matter. Recently discovered topological semimetals are believed to host a nontrivial $\pi$ Berry phase to induce a phase shift of $\pm 1/8$ in the quantum oscillation ($+$ for hole and $-$ for electron carriers). We theoretically study the Shubnikov-de Haas oscillation of Weyl and Dirac semimetals, taking into account their topological nature and inter-Landau band scattering. For a Weyl semimetal with broken time-reversal symmetry, the phase shift is found to change nonmonotonically and go beyond known values of $\pm 1/8$ and $\pm 5/8$. For a Dirac semimetal or paramagnetic Weyl semimetal, time-reversal symmetry leads to a discrete phase shift of $\pm 1/8$ or $\pm 5/8$, as a function of the Fermi energy. Different from the previous works, we find that the topological band inversion can lead to beating patterns in the absence of Zeeman splitting. We also find the resistivity peaks should be assigned integers in the Landau index plot. Our findings may account for recent experiments in Cd$_2$As$_3$ and should be helpful for exploring the Berry phase in various 3D systems.
\end{abstract}


\maketitle

The Shubnikov-de Haas oscillation of resistance in a metal arises from the Landau quantization of electronic states under strong magnetic fields. The oscillation can be described by the Lifshitz-Kosevich formula \cite{Shoenberg84book} $\cos[2\pi(F/B+\phi)]$, where $B$ is the magnitude of magnetic field, and the oscillation frequency $F$ and phase shift $\phi$ can provide valuable information about the Fermi surface topography of materials. It is widely believed that an energy band with linear dispersion carries an extra $\pi$ Berry phase \cite{Miktik99prl,Xiao10rmp}, leading to phase shifts of $\phi=0$ and $\pm 1/8$ in 2D and 3D, respectively, compared with $\pm 1/2$ and $\pm 5/8$ for parabolic energy bands without the Berry phase ($+$ for hole and $- $ for electron carriers).
Topological semimetals \cite{Volovik03book,Wan11prb,Xu11prl,Burkov11prl,Yang11prb} provide a new platform to study the nontrivial Berry phase in 3D. They have linear dispersion near the Weyl nodes at which the conduction and valence bands touch. The Weyl nodes host monopoles connected by Fermi arcs, and have been discovered in the Dirac semimetals Na$_3$Bi \cite{Wang12prb,Liu14sci,Xu15sci} and Cd$_3$As$_2$ \cite{Wang13prb,Liu14natmat,Neupane14nc,Borisenko14prl,Yi14srep,ZhangC15arXiv,LiCZ15nc,LiH16nc}, and the Weyl semimetals TaAs family  \cite{Huang15nc,Weng15prx,Lv15prx,Xu15sci-TaAs,Yang15np,ZhangCL16nc,HuangXC15prx,Yang15arXiv, Xu15np-NbAs,Xu16nc-TaP} and YbMnBi$_2$ \cite{Borisenko15arXiv}.

Exploring the $\pi$ Berry phase in 3D semimetals remains difficult \cite{Murakawa13sci,He14prl,Novak15prb,Zhao15prx,Du16scpma,YangXJ15arXiv-NbAs,WangZ16prb,Xiong15sci,Cao15nc,ZhangCL15arXiv-TaAs,Narayanan15prl,Park11prl,Xiang15prb,Tafti16np,Luo15prb,Dai16prbrc}.
To extract the phase shift, the Landau indices, i.e., where $F/B+\phi$ takes integers $n$, need to be identified first from the magnetoresistivity. A plot of $n$ vs $1/B$ then extrapolates to the phase shift on the $n$ axis. However, the first step in 3D is highly nontrivial. In 3D, a magnetic field quantizes the energy spectrum into a set of 1D bands of Landau levels. There may be multiple Landau bands on the Fermi surface and scattering among them. This situation never occurs for discrete Landau levels in 2D. It is not intuitive to determine the Landau indices in 3D without a sophisticated theoretical analysis of the resistivity of the Landau bands. Both the resistivity peaks \cite{Murakawa13sci,Zhao15prx,He14prl,Zhao15prx,Du16scpma,YangXJ15arXiv-NbAs,WangZ16prb,Cao15nc} and valleys \cite{Narayanan15prl,Qu10sci,Park11prl,Luo15prb} have been used to identify the Landau indices in different experiments. The treatments can introduce a system error of $\pi$, comparable with the $\pi$ Berry phase under quest, and partially lead to a wide range of the phase shifts away from the anticipated $\pm 1/8$ in the experiments (see Sec. S1 of Ref. \cite{Supp}).

In this Letter, we calculate the resistivity in both longitudinal and perpendicular magnetic fields for topological Weyl and Dirac semimetals. We clarify explicitly that the resistivity peaks appear near Landau band edges and correspond to integer Landau indices. For time-reversal symmetry broken Weyl semimetals, we find that the phase shift can go beyond known values of $\pm 1/8$ or $\pm 5/8$ and nonmonotonically approach a wide range between $\pm 7/8$ and $\pm 9/8$ near the Lifshitz point, and these values may be misinterpreted as $\pm 1/8$ in experiments. For Dirac semimetals or Weyl semimetals with time-reversal symmetry, the combined phase shift takes the discrete values of either $\pm 1/8$ or $\pm 5/8$. Moreover, a new beating pattern, due to the topological band inversion rather than Zeeman splitting, is found. Our findings may explain the positive phase shifts of electron carriers in recent experiments, and should be helpful for experiments involving the Berry phase and monopole physics in various 3D systems.

\begin{figure}[htpb]
\centering
\includegraphics[width=0.4 \textwidth]{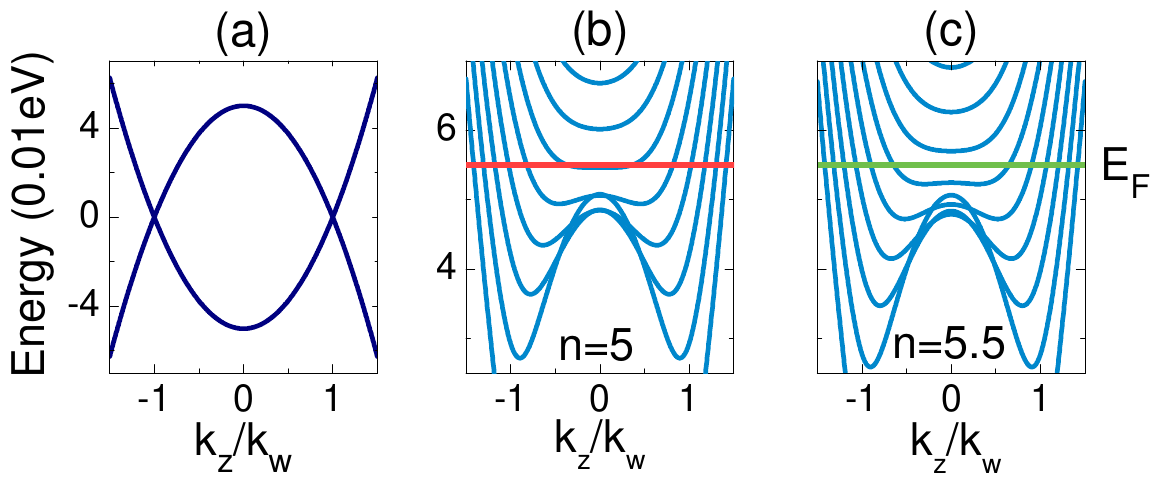}
\includegraphics[width=0.4 \textwidth]{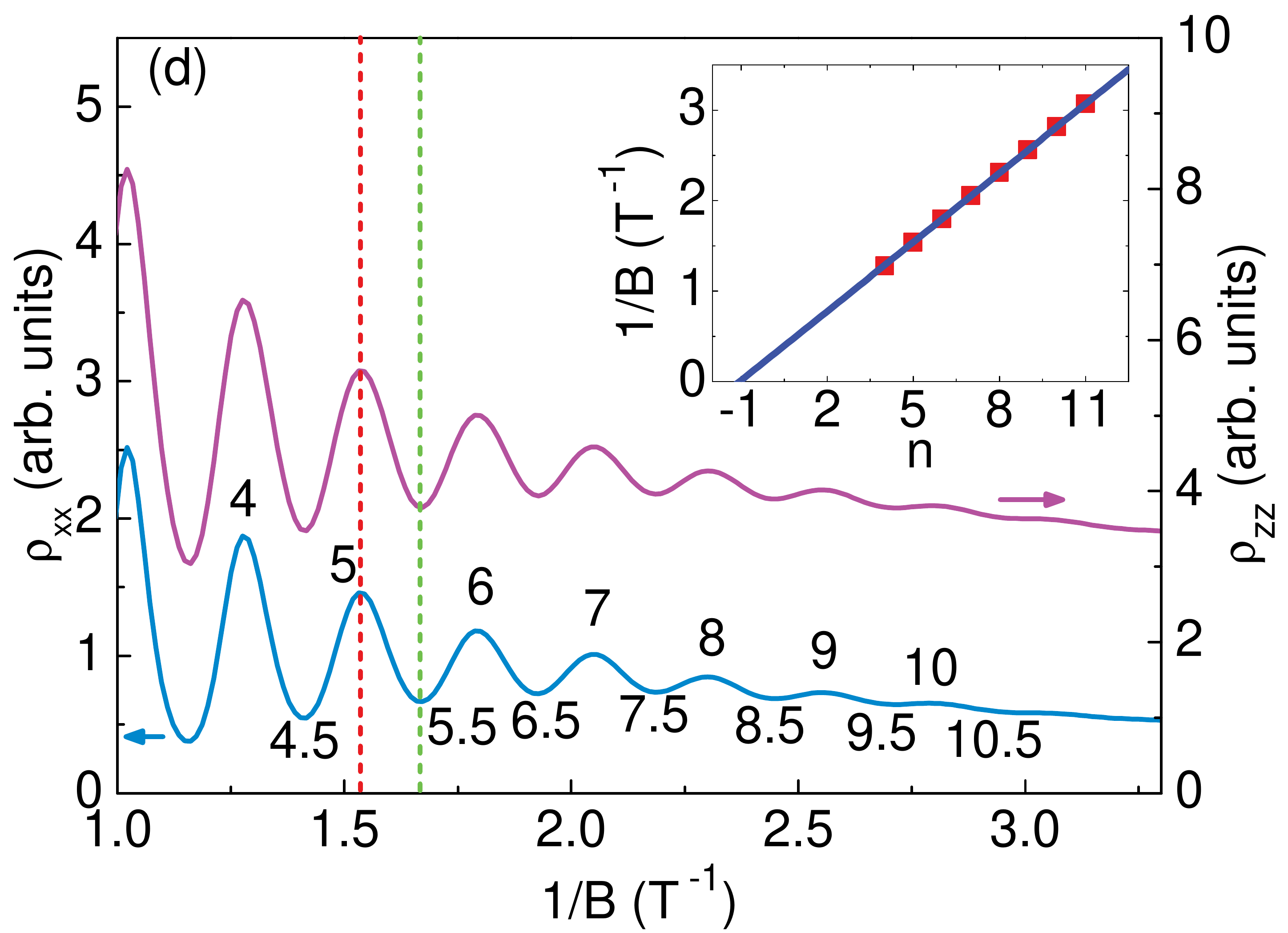}
\caption{(Color online) (a) The conduction and valence bands of the Weyl semimetal as a function of $k_z$ at $k_x=k_y=0$. [(b) and (c)] In the $z$-direction magnetic field, the Landau bands when $n=5$ and $5.5$ [see dash lines in (d)]. (d) An example of the numerically calculated resistivities $\rho_{xx}$ and $\rho_{zz}$ as functions of $1/B$. Inset: The Landau index plot and linear fitting (line) using $n=F/B+\phi$ to the peaks in $\rho_{xx}$ and $\rho_{zz}$. In this case, $F= 3.927\pm 0.003$ and $\phi=-1.052\pm 0.007$. The parameters are $k_w=0.1$ nm$^{-1}$, $A=0.5$ eV nm, $M=5$ eV nm$^2$, and the Fermi energy $E_F=0.055$ eV.}\label{fig:fitting}
\end{figure}

\emph{Model.} - We start from a two-node Hamiltonian for a Weyl semimetal \cite{Shen12book,Okugawa14prb,Lu15Weyl-shortrange}
\begin{equation}\label{Ham}
\mathcal H =A(k_x\sigma_x+k_y\sigma_y)+M (  k_w^2-\mathbf{k}^2)\sigma_z,
\end{equation}
where $ (\sigma_x,\sigma_y,\sigma_z)$ are the Pauli matrices, the wave vector $\mathbf{k}=(k_x,k_y,k_z)$, and $A$, $M$, and $k_w$ are model parameters.
The energy dispersion of the model is $E_{\pm }^{\mathbf{k}}=\pm [M^2(k_w^2-\mathbf{k}^2)^2 +A^2(k_x^2+k_y^2)]^{1/2}$, with $\pm$ for the conduction and valence bands, respectively. The model hosts two Weyl nodes at $(0,0,\pm k_w)$ [Fig. \ref{fig:fitting}(a)], and has been demonstrated to carry all of the topological semimetal properties \cite{Lu15Weyl-shortrange}. In particular, the Fermi arcs, i.e., the $k_z$-dependent topological edge states, can be solved analytically from the model with an open boundary condition \cite{ZhangSB16njp}, in contrast to the $k\cdot \sigma$ model \cite{Phillip14epjb,Gorbar14prb,Lu15Weyl-localization}. The topological properties of the model arise from the $\sigma_z$ term \cite{Lu10prb}, with which the model can smoothly change from linear dispersion near the Weyl nodes to parabolic dispersion at high Fermi energies.

\emph{Quantum oscillation in linear and parabolic limits.} - In the presence of a $z$-direction magnetic field $ B $, the energy spectrum splits into a series of 1D bands of Landau levels \cite{Lu15Weyl-shortrange,ZhangSB16njp} [see Figs. \ref{fig:fitting}(b)-\ref{fig:fitting}(c)], which give rise to the quantum oscillation. We focus on the bulk states, as the oscillation via surface states requires ultrathin films \cite{Potter14nc} and can be ignored in the work.
We calculate the resistivity in two direction configurations following linear response theory \cite{Charbonneau82jmp,Vasilopoulos84jmp, WangCM12prb,WangCM15prb} (see Sec. S2 of Ref. \cite{Supp} for the calculation details). In the longitudinal configuration, resistivity is measured along the $z$ direction (denoted as $\rho_{zz}$), and in the transverse configuration resistivity is measured along the $x$ direction ($\rho_{xx}$). The magnetoresistivity in the linear and parabolic dispersion limits can be found analytically to take the general form
\begin{equation}
(\rho-\rho_0)/\rho_0 = \mathcal{C} \exp(-\lambda_D)\cos\left[2\pi\left(F/B+\phi\right)\right] \label{rho},
\end{equation}
where subscripts $xx$ and $zz$ are suppressed for simplicity, $\rho_0$ is the zero-field resistivity, $\lambda_D$ is the Dingle factor, and $\mathcal{C}$ is a constant coefficient. The analytic expressions for the frequency $F$ and phase shift $\phi$ are listed in Table \ref{tab:two-limits} for the two limits. We can analytically obtain the expected $-1/8$ in the linear limit and $-5/8$ in the parabolic limit for electron carriers.
Note that the frequency in the linear limit depends not on the effective mass, but on $A$ in the velocity term.

\begin{table}[htb]
    \centering
\caption{The analytical expressions for the frequency $F$ and phase shift $\phi$ in the resistivity formula Eq. \eqref{rho} in the linear and parabolic dispersion limits for electron carriers. We define $E_F'\equiv E_F+ Mk_w^2 $. }\label{tab:two-limits}
\begin{ruledtabular}
\begin{tabular}{ccccc}
                           & \multicolumn{2}{c}{Longitudinal $\rho_{zz}$} & \multicolumn{2}{c}{Transverse $\rho_{xx}$} \\
                     &Parabolic                 &Linear       &Parabolic     &Linear\\ \hline
\rule{0pt}{0.65cm}$F$      & $\hbar E_F'/2eM$       & $\hbar E_F^2/2eA^2$           &  $\hbar E_F'/2eM$            & $\hbar E_F^2/2eA^2$      \\
\rule{0pt}{0.65cm}$\phi$             & -5/8            & -1/8 & -5/8 & -1/8\\
\end{tabular}
\end{ruledtabular}
\end{table}

\begin{figure}
\centering
\includegraphics[width=0.38\textwidth]{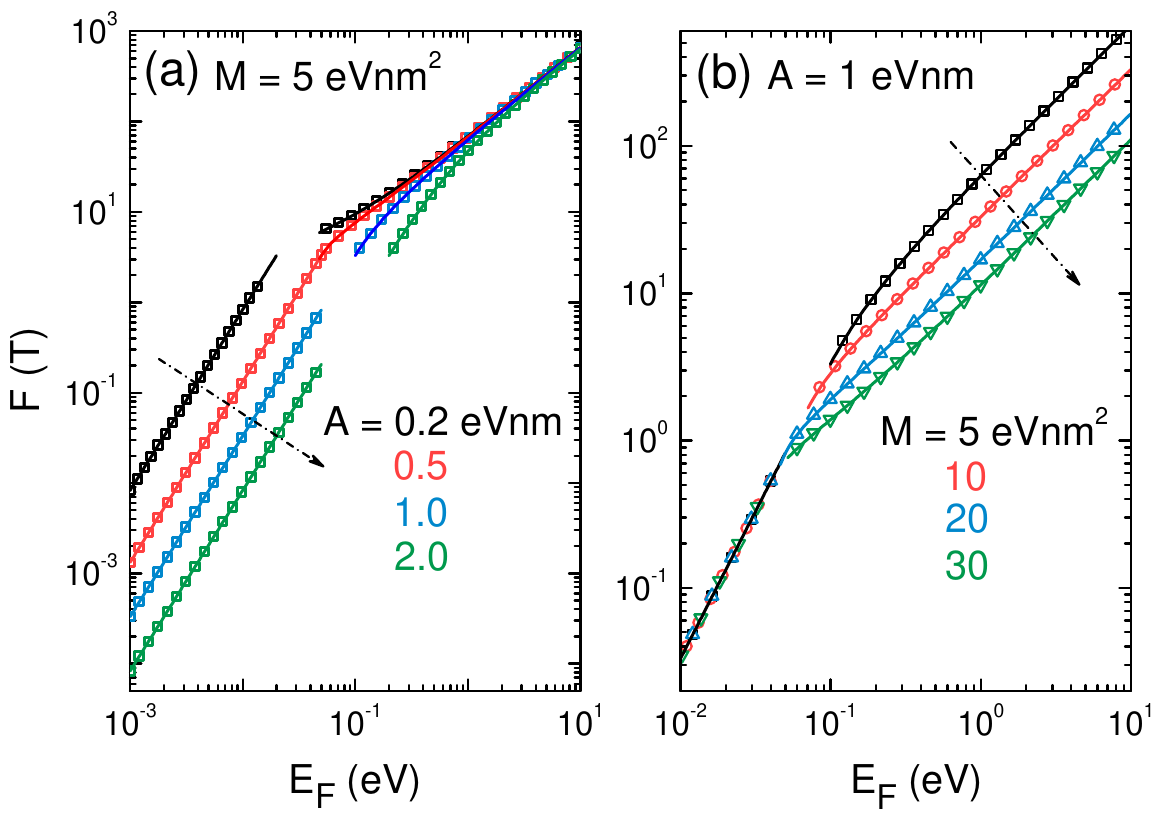}
\includegraphics[width=0.4\textwidth]{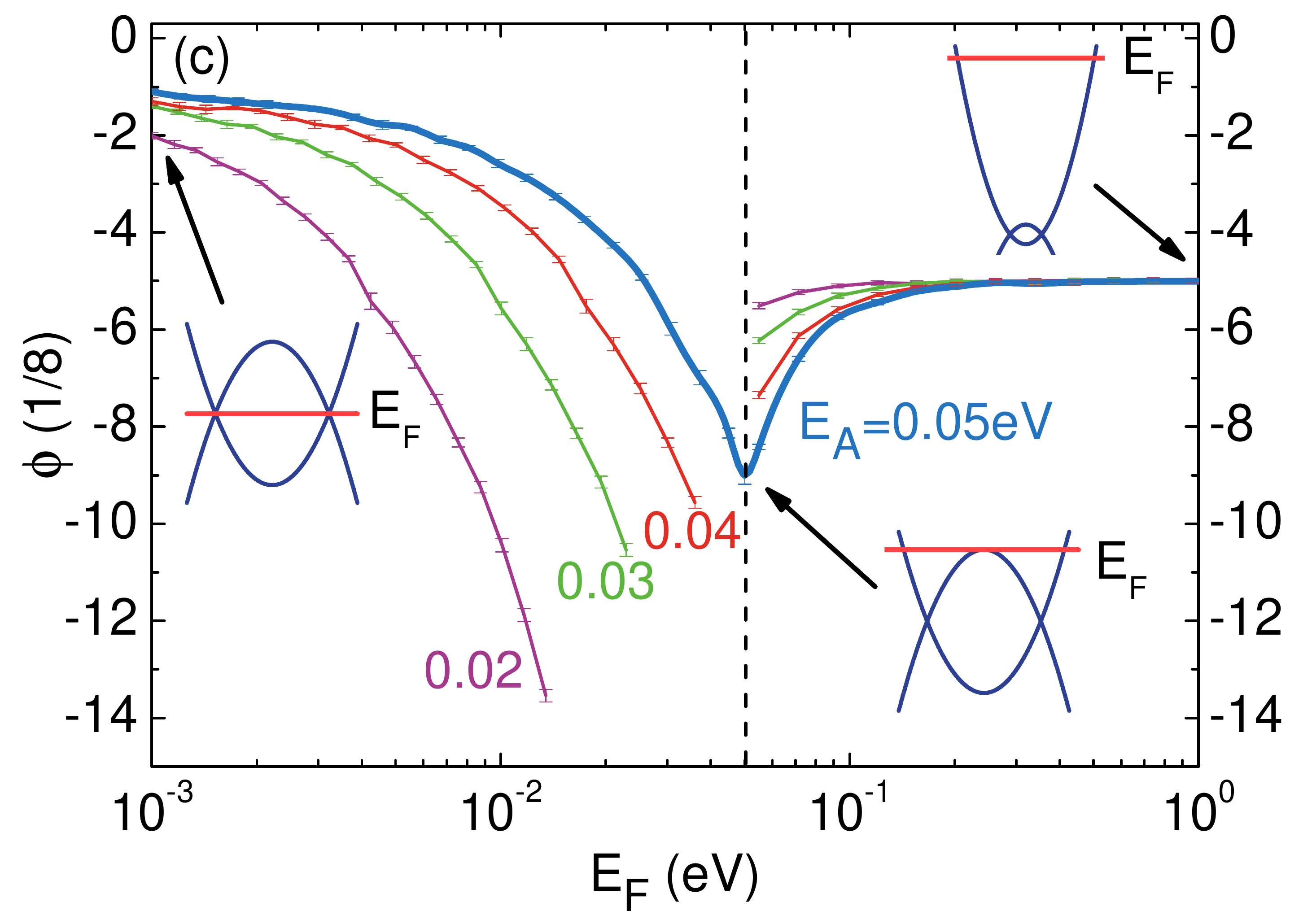}
\caption{(Color online) For the Weyl semimetal with broken time-reversal symmetry. (a) The frequency $F$ obtained numerically (scatters) and analytically (solid curves) vs the Fermi energy $E_F$ for (a) different $A$ at a fixed $M$; and (b) for different $M$ at a fixed $A$. (c) The phase shift $\phi$ vs $E_F$ for different $E_A=Ak_w$ and a fixed $E_M=Mk_w^2=0.05$ eV. The curves break because $F$ and $\phi$ cannot be fitted when beating patterns form. The insets indicate the location of Fermi energy. The vertical dashed lines mark the Lifshitz point. $k_w=0.1$ nm$^{-1}$ throughout the work.}\label{fig:phase-shift}
\end{figure}

\emph{Resistivity peaks and integer Landau indices.} -
In experiments, due to sophisticated data patterns, the oscillation may not be well fitted by the Lifshitz-Kosevich form in Eq. (\ref{rho}). Instead, the peak or valley positions on the $B$ axis are assigned integer Landau indices $n$, then $\phi$ and $F$ can be fitted from a plot of $n$ and $1/B$ [see inset of Fig. \ref{fig:fitting}(d)]. However, whether the peaks \cite{Murakawa13sci,Zhao15prx,He14prl,Zhao15prx,Du16scpma,YangXJ15arXiv-NbAs,WangZ16prb,Cao15nc} or valleys \cite{Narayanan15prl,Qu10sci,Park11prl,Luo15prb} should be assigned indices is still in debate.
Our results explicitly clarify that the resistivity peaks of both $\rho_{xx}$ and $\rho_{zz}$ appear near Landau band edges and correspond to integer Landau indices. As shown in Fig. \ref{fig:fitting}, peak 5 in (d) appears when the Fermi energy is close to the band edge of the 5th Landau band [Fig. \ref{fig:fitting}(b)], valley 5.5 appears when the Fermi energy lies somewhere between the 5th and 6th bands [Fig. \ref{fig:fitting}(c)]. The numerical results using the peaks as integers in the Landau index plot are shown in Fig. \ref{fig:phase-shift}. As shown in Fig. \ref{fig:phase-shift}(c), in the limits $E_F\rightarrow 0$ and $\infty$, the numerical fitting can recover the analytic results of -1/8 and -5/8 phase shifts, respectively.

Why both $\rho_{zz}$ and $\rho_{xx}$ show peaks near the band edges can be explained as follows. In theory, the resistivity components are evaluated from the conductivity components \cite{Datta1997,Vasko06book}. In the longitudinal configuration, the resistivity $\rho_{zz}$=$1/\sigma_{zz}$, where $\sigma_{zz}$ is the conductivity along the $z$ direction. Near the band edges, because of vanishing velocities, the conductivity $\sigma_{zz}$ shows valleys, so $\rho_{zz}$ shows peaks.
In the transverse configuration, $\rho_{xx}=\sigma_{yy}/(\sigma_{yy}^2+\sigma_{xy}^2)$, and the longitudinal and field-induced Hall conductivities are found as (see Sec. S3 of Ref. \cite{Supp} for the calculation)
\begin{eqnarray}\label{sigma-yx-delta}
\sigma_{yy}= \frac{\sigma_0(1+\delta)}{1+(\mu B)^2},\ \  \sigma_{yx} = \frac{ \mu B \sigma_0 }{1+(\mu B)^2}\left[1-\frac{\delta}{(\mu B)^2} \right],
\end{eqnarray}
where $\sigma_0$ is the zero-field conductivity and $\delta\ll 1$ represents the oscillation part. The $\delta$ term in $\sigma_{yx}$ is from the disorder scattering and was seldom considered before. A consequence of the $\delta$ term in $\sigma_{yx}$ is that $\rho_{xx} \approx  (1+\delta)/\sigma_0 $,
up to the leading order of $\delta$. As both $\rho_{xx}$ and $\sigma_{yy}$ are proportional to $1+\delta$, their peaks are aligned for the arbitrary ratio of $\sigma_{yx}$ to $\sigma_{yy}$ (but the oscillation is too weak to be observed when $\sigma_{yx}\ll \sigma_{yy}$). This is a new finding as a result of the disorder scattering $\delta$ term in the Hall conductance. Meanwhile, the $\sigma_{zz}$ valleys are aligned with the $\sigma_{yy}$ peaks, because $\sigma_{zz}$ originates from diffusion and is proportional to the scattering times, while $\sigma_{yy}$ results from hopping (i.e., off-diagonal velocities and scattering times that couple different Landau bands) and is inversely proportional to the scattering times \cite{Abrikosov98prb,Lu15Weyl-shortrange,Vasilopoulos84jmp}. Stronger scattering (i.e., shorter scattering times) can suppress diffusion but enhance hopping. In summary, the peak positions satisfy the relation
$\rho_{zz} \sim \sigma_{zz}^{-1} \sim  \sigma_{yy} \sim \rho_{xx} $, so both $\rho_{zz}$ and $\rho_{xx}$ show peaks near the Landau band edges and share the same phase shift.

\emph{Anomalous phase shift near the Lifshitz point.} -
For a Fermi energy between the linear and parabolic dispersion limits, the phase shift is expected to change from one limit to the other. However, we find the crossover can be nonmonotonic. We numerically calculate the frequency and phase shift by fitting the Landau index plot [see inset of Fig. \ref{fig:fitting}(d)] for arbitrary Fermi energy $E_F$.
Figure \ref{fig:phase-shift} shows the numerical results for $F$ and $\phi$. In Figs. \ref{fig:phase-shift} (a) and \ref{fig:phase-shift}(b), the comparison between the analytical [see Eq. (\ref{Bfeq}) and its vicinity] and numerical results for $F$ justifies our numerical scheme. Note that $F$ converges because it depends only on $A$ near the Weyl nodes and only on $M$ at higher Fermi energies.
In Fig. \ref{fig:phase-shift}(c), the numerical results recover the analytical $\phi=-1/8$ in the linear limit ($E_F\rightarrow 0$) and $-5/8$ in the parabolic limit ($E_F\rightarrow \infty $).
For convenience, we define two energy parameters,
\begin{equation}
E_A= A k_w,\ \ \ E_M=Mk_w^2.
\end{equation}
For $E_M\neq E_A$, the $\phi$-$E_F$ curves break due to the formation of beating patterns, which we discuss later. In Fig. \ref{fig:phase-shift} (c), when $E_A<E_M$, the phase shift does not monotonically transit from -1/8 to -5/8, but drops below $-5/8$ in an intermediate regime around the so-called Lifshitz transition point (at which $E_F=E_M$). In either the linear or parabolic limit, the energy spectrum is a simple function of $k_z^2$, and an integral of $k_z$ gives the extra $\pm 1/8$ phase compared with that in 2D.
Away from the two limits, this simple $k_z^2$ dependence is violated, which is probably the reason for the anomalous phase shift. We can analytically show the phase shift of $-9/8$ at the Lifshitz point when $E_M=E_A$ (see Sec. S4 of Ref. \cite{Supp}), consistent with that in Fig.~\ref{fig:phase-shift} (c).
This value is equivalent to $-1/8$, which is
usually believed to arise from the $\pi$ Berry phase when the Fermi sphere encloses single Weyl nodes. However, in this case, the Fermi sphere encloses two Weyl nodes with a Fermi energy at the Lifshitz point.
When $E_A> E_M $, there is no nonmonotonicity in $\phi-E_F$.

\begin{figure}
\centering
\includegraphics[width=0.38\textwidth]{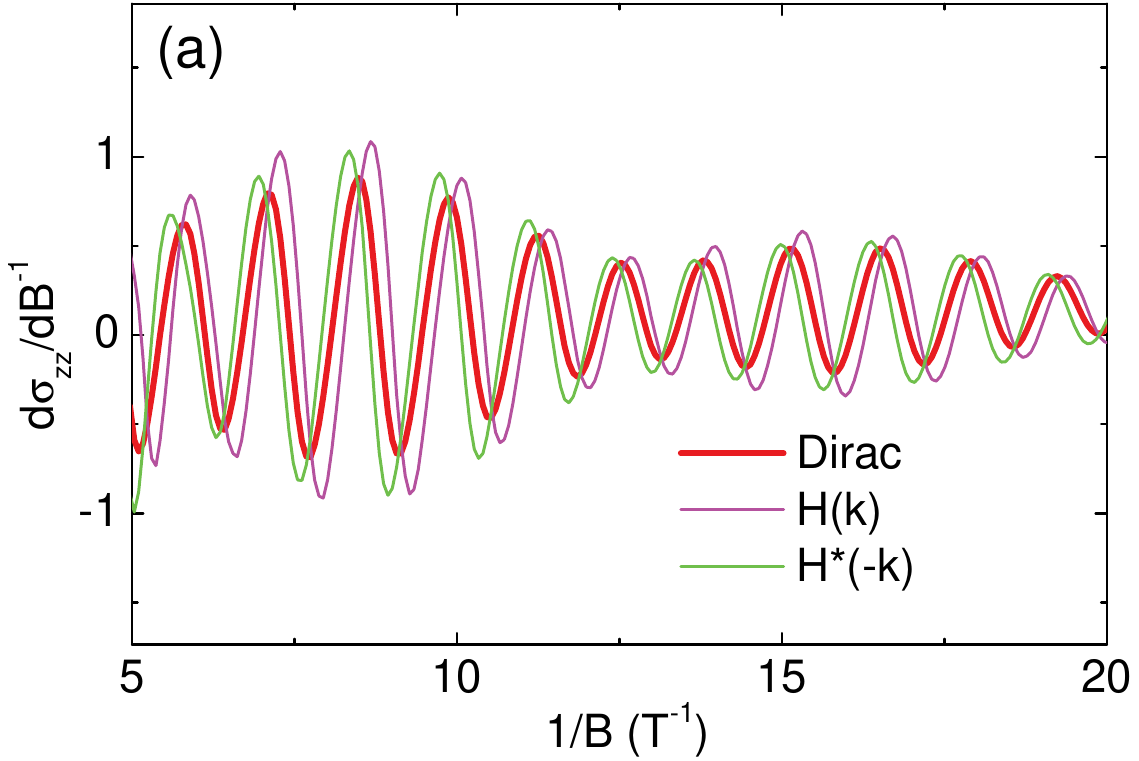}
\includegraphics[width=0.39\textwidth]{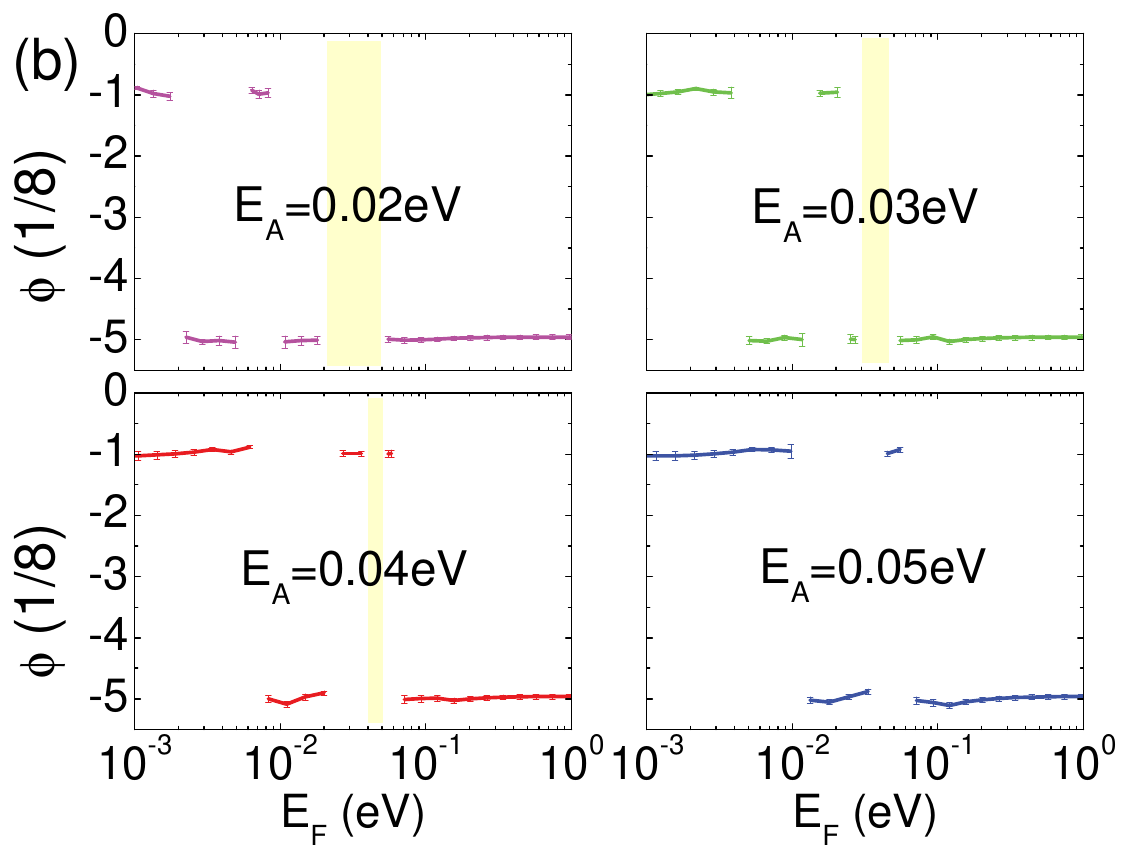}
\caption{(Color online) For the Dirac semimetal or Weyl semimetal with time-reversal symmetry. (a) $d\sigma_{zz}/d(1/B)$ as a function of $1/B$ for the Dirac semimetal and its Weyl components [$\mathcal{H}(\mathbf{k})$ and $\mathcal{H}^*(-\mathbf{k})$]. The parameters: $E_F=0.0954$ eV, $ k_w =0.1$/nm, $A=2\,{\rm eVnm}$, and $M=5$ eV nm$^2$, so that $E_F<{\rm min}(E_A,E_M)$. (b) Each panel is the same as Fig. \ref{fig:phase-shift}(c) but for the Dirac semimetal. The data break because $\phi$ cannot be fitted when beating patterns form (highlighted area).}\label{fig:phase-shift-Dirac}
\end{figure}

A Weyl semimetal and its time-reversal partner can form a Dirac semimetal, whose model can be built by $\mathcal{H}(\mathbf{k})$ in Eq. (\ref{Ham}) and its time-reversal partner $\mathcal{H}^*(-\mathbf{k})$, where the asterisk refers to a complex conjugate. This model can also serve as a building block for Weyl semimetals that respect time-reversal symmetry but break inversion symmetry \cite{Huang15nc,Weng15prx,Lv15prx,Xu15sci-TaAs,Yang15np,ZhangCL16nc,HuangXC15prx,Yang15arXiv, Xu15np-NbAs,Xu16nc-TaP}. For this case, there is no anomalous Hall effect. The change of phase shift of $\mathcal{H}^*(-\mathbf{k})$ is opposite to that of $\mathcal{H}(\mathbf{k})$, and the two give rise to a {\it combined} phase shift. If we describe the oscillation of $\mathcal{H}(\mathbf{k})$ by $\cos[2\pi(F/B+\alpha -1/8)]$, then that of $\mathcal{H}^*(-\mathbf{k})$ is $\cos[2\pi(F/B-\alpha-1/8)]$, and the oscillation of the Dirac semimetal behaves like $\cos(2\pi\alpha ) \cos [2\pi (F/B-1/8)]$. According to Fig. \ref{fig:phase-shift}(c), $\alpha$ can vary over 1, so $\cos(2\pi\alpha)$ may be negative, giving an extra $\pi$ phase shift. In this case, the combined phase shift of the Dirac semimetal may take two discrete values, $-1/8$ when $\alpha\in [0,1/4]$ and $[3/4,1]$ or $-5/8$ when $\alpha\in [1/4,3/4]$, as shown in Fig. \ref{fig:phase-shift-Dirac} (b).
The combined phase shift tends to be $-1/8$ near the Weyl nodes and $-5/8$ at higher Fermi energies. Near the Lifshitz point, the combined phase shift may jump between the two values.
The scattering between $\mathcal{H}(\mathbf{k})$ and $\mathcal{H}^*(-\mathbf{k})$ is fully considered in the calculation and adheres to the preceding argument.

\begin{table}[htbp]
    \centering
\caption{The phase shift $\phi_{\text{exp}}$ extracted from the experiments on Cd$_3$As$_2$. According to the theory in this work, if peaks from two Weyl components can be distinguished, the phase shift should be $ \phi_{\text{Weyl}}=\phi_{\text{exp}}-1$; otherwise, the phase shift should be $\phi_{\text{Dirac}}=-5/8$ according to
Fig. \ref{fig:phase-shift-Dirac}(b).  Compared with Eq. (\ref{rho}), the formula used in Refs. \cite{He14prl} and \cite{Narayanan15prl} has an extra $1/2$, which has been subtracted from $\phi_{\text{exp}}$. }\label{tab:eps-phase}
\begin{ruledtabular}
\begin{tabular}{cccc}
    Ref.      & $\phi_{\text{exp}}$ &  $\phi_{\text{Weyl}}$ & $\phi_{\text{Dirac}}$\\
 \hline
 \cite{He14prl}   & 0.06 $\sim$ 0.08& -0.94 $\sim$ -0.92 &  -5/8  \\
 \cite{Zhao15prx}  &0.11 $\sim$ 0.38&-0.89 $\sim$ -0.62 &  -5/8\\
\cite{Narayanan15prl}  & 0.04\footnote{Read from Fig 2(d) of Ref. \cite{Narayanan15prl}.} &-0.96 &  -5/8
\end{tabular}
\end{ruledtabular}
\end{table}

The anomalous phase shift in Figs. \ref{fig:phase-shift} and \ref{fig:phase-shift-Dirac} probably has been observed in the experiments. The electron and hole carriers are supposed to yield negative and positive phase shifts, respectively \cite{Murakawa13sci}. However, the phase shift in the Dirac semimetal Cd$_3$As$_2$ experiments are found to take positive values for electron carriers \cite{He14prl,Narayanan15prl,Zhao15prx}. One possible explanation is that the actual values of the phase shift in the experiments are around $-7/8$ to $-5/8$, and hence look like $1/8$ to $3/8$ because of the $2\pi$ periodicity. According to our numerical results, the combined phase shift takes these values from around the Lifshitz point to higher Fermi energies, which is also consistent with the carrier density in the experiments. In Table \ref{tab:eps-phase}, we suggest the counterparts for the experimental values of the phase shift. Nevertheless, a comparison with the TaAs family is difficult, because there are too many bands on the Fermi surface.

\emph{Beating pattern from topological band inversion.} -
Figure \ref{fig:phase-shift-Dirac} (a) also shows that the Dirac semimetal and each of its Weyl components develop beating patterns. They are not from the Zeeman splitting, but inherited from the band inversion of the Weyl semimetal [see Fig. \ref{fig:fitting}(a)]. Consequently, some Landau bands have more than one extreme point [see Figs. \ref{fig:fitting} (b)-\ref{fig:fitting}(c)]. We can also show that each extreme point gives rise to a resistivity peak. Then, for each Weyl component, the oscillation may have two frequencies and develop beating patterns. We find the frequency analytically. For $E_F$ below $E_A$ and $E_M$,
$F= F_{0} \equiv E_F^2\hbar/2A^2e $; for $E_F$ between $E_A$ and $E_M$,
\begin{eqnarray}\label{Bfeq}
F=\left\{\begin{array}{lc}
F_{+}\ \text{and}\ F_{0},& E_M <E_F<E_A,\\
F_{+}\ \text{and}\ F_{-},&E_A<E_F< E_M; \\
\end{array}\right.
\end{eqnarray}
and for $E_F$ above $E_A$ and $E_M$, $F=F_+$, where
$ F_{\pm }  = (\hbar/e)(E_F^2-E_M^2)/\{  A^2-2 ME_M  \pm [(2ME_F)^2-( 2ME_A)^2+A^4 ]^{1/2}\}$.
Equation (\ref{Bfeq}) indicates that the beating pattern forms only when the Fermi energy $E_F$ is between $E_M$ and $E_A$.
Here, the beating pattern arises because of the topological nature of the semimetal, different from the Zeeman splitting \cite{ZhangCL15arXiv-TaAs,Hu16srep,Cao15nc}, nested Fermi surfaces \cite{Zhao15prx}, and orbital quantum interference \cite{Xiong15arXiv}.

\begin{acknowledgments}
We thank P. Vasilopoulos for valuable discussions. This work was supported by the National Key R \& D Program (Grant No. 2016YFA0301700), the National Science Foundation of China (Grant Nos. 11474005 and 11574127) and the Research Grant Council, University
Grants Committee, Hong Kong (Grant No. 17303714).
\end{acknowledgments}

%

%

\newpage

\includepdf[pages=1]{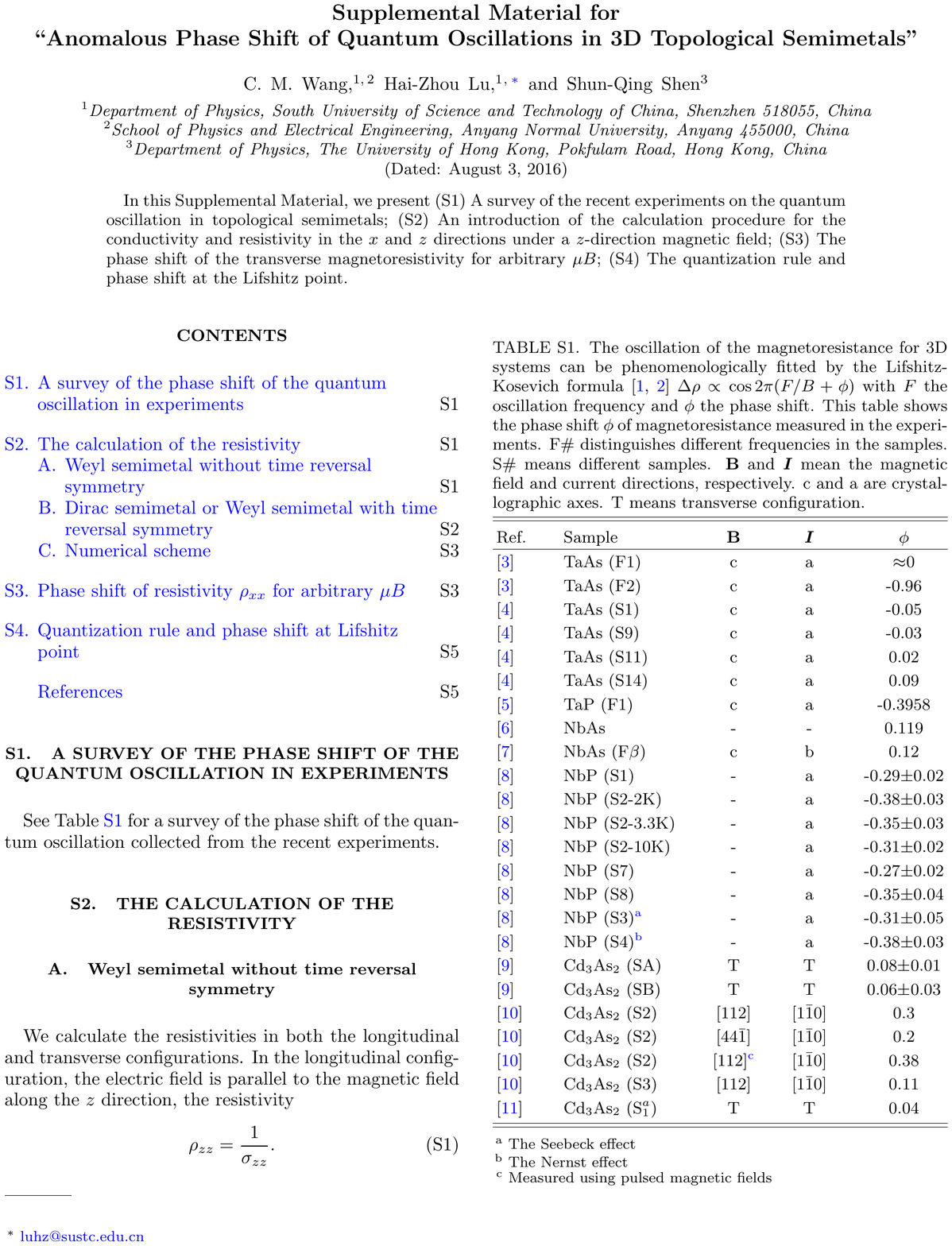}

\ \\

\newpage

\includepdf[pages=2]{Weylosci_supp_20160721.pdf}

\ \\

\newpage

\includepdf[pages=3]{Weylosci_supp_20160721.pdf}

\ \\

\newpage

\includepdf[pages=4]{Weylosci_supp_20160721.pdf}

\ \\

\newpage

\includepdf[pages=5]{Weylosci_supp_20160721.pdf}

\end{document}